\documentclass[10pt,journal,compsoc]{IEEEtran}

\ifCLASSOPTIONcompsoc
  \usepackage[nocompress]{cite}
\else
  \usepackage{cite}
\fi

\usepackage{graphicx}
\usepackage{subcaption}
\usepackage{booktabs}
\usepackage{multirow}
\usepackage{hyperref}
\usepackage{amsmath}
\usepackage{afterpage}
\begin{document}

\title{An Integrated Framework for Process Discovery Algorithm Evaluation}

\author{Toon~Jouck,
        Alfredo~Bolt,
				Beno\^it~Depaire,
				Massimiliano~de~Leoni,
        and~Wil~M.~P.~van~der~Aalst
\IEEEcompsocitemizethanks{\IEEEcompsocthanksitem T. Jouck and Beno\^it Depaire are with the Faculty of Business Economics, UHasselt - Hasselt University, 3590 Diepenbeek,
Belgium.\protect\\
E-mail: toon.jouck, benoit.depaire@uhasselt.be.
\IEEEcompsocthanksitem A. Bolt, M. de Leoni and W. van der Aalst are with the Department of Mathematics and Computer Science, Eindhoven University of Technology, Eindhoven, The Netherlands.\protect\\
E-mail: a.bolt, m.d.leoni, w.m.p.v.d.aalst@tue.nl.}}


\markboth{IEEE Transactions on Knowledge and Data Engineering}%
{Jouck \MakeLowercase{\textit{et al.}}: Integrated Framework for Process Discovery Algorithm Evaluation}

\IEEEtitleabstractindextext{%
\begin{abstract}
Process mining offers techniques to exploit event data by providing insights and recommendations to improve business processes. The growing amount of algorithms for process discovery has raised the question of which algorithms perform best on a given event log. Current evaluation frameworks for empirically evaluating discovery techniques depend on the notation used (behavioral identical models may give different results) and cannot provide more general statements about populations of models. Therefore, this paper proposes a new integrated evaluation framework that uses a classification approach to make it modeling notation independent. Furthermore, it is founded on experimental design to ensure the generalization of results. It supports two main evaluation objectives: benchmarking process discovery algorithms and sensitivity analysis, i.e. studying the effect of model and log characteristics on a discovery algorithm's accuracy. The framework is designed as a scientific workflow which enables automated, extendable and shareable evaluation experiments. An extensive experiment including four discovery algorithms and six control-flow characteristics validates the relevance and flexibility of the framework. Ultimately, the paper aims to advance the state-of-the-art for evaluating process discovery techniques.
\end{abstract}

}

\maketitle

\IEEEdisplaynontitleabstractindextext

\IEEEpeerreviewmaketitle

\IEEEraisesectionheading{\section{Introduction}\label{sec:introduction}}

\IEEEPARstart{T}{oday's} information systems store large amounts of data about the business processes within organizations. This leads to the challenge of extracting value and information out of these event data.
Process mining is a discipline that sits between data mining and process modeling and analysis and, hence, can be considered as the linking-pin between data science and process science~\cite{van_der_aalst_process_2016}.
The idea of process mining is to discover, monitor and improve the processes by extracting knowledge from the data that are stored in information systems about how these systems are used to carry out processes.
Differently from a-priori analysis, the focus is not on the assumed processes but on real processes in the way that they are executed. Therefore, the starting point of process mining is an event log, which is analyzed to extract useful insights and recurrent patterns about how processes are executed within organizations.

The lion's share of attention within process mining was received by process discovery, which aims to discover a process model from event logs. This resulted in dozens of new discovery algorithms (see, e.g.,\cite{agrawal_mining_1998,shao_efficient_2008,gaaloul_discovering_2004} and for an overview see~\cite{van_der_aalst_process_2016,de_weerdt_multi-dimensional_2012}). Researchers aim to improve the quality of the mined models to adequately represent the behavior observed in the event logs. Typically, the quality is measured as the fitness between the event log and the mined model. A good model allows for the behavior seen in the event log. Fitness alone is not sufficient, also a proper balance between \emph{overfitting} and \emph{underfitting} is required~\cite{van_der_aalst_process_2016}. A process model is overfitting (the event log) if it is too restrictive, disallowing behavior which is part of the underlying process. This typically occurs when the model only allows for the behavior recorded in the event log. Conversely, it is underfitting (the reality) if it is not restrictive enough, allowing behavior which is unlikely to be part of the underlying process. This typically occurs if it overgeneralizes the observed behavior in the event log.

The abundance of discovery algorithms has made it increasingly important to develop evaluation frameworks that can compare the efficiency of these discovery techniques, especially in terms of balancing between overfitting and underfitting. As detailed in Section~\ref{sec:related_work}, several comparison frameworks have already been proposed in literature. Unfortunately, these frameworks are characterized by at least one of the following three major limitations:
\begin{enumerate}
  \item They are not independent from the modeling notation in which the discovered models are represented, e.g. two behaviorally equivalent models may have very different precision scores, or quality can only be measured after a conversion that does not preserve the behavior precisely. This restricts the framework to a comparison of the algorithms that generate models in one specific notation.
  \item The evaluation results are based on real event logs and cannot be generalized as the population of processes from which they originate is unknown. Processes come from different populations depending on the type of behavior allowed. Processes may have different behavioral characteristics, with parts that can repeat, with mutually-exclusive and parallel branches, with non-local dependencies and so on. Also, these characteristics can be more or less predominant in a process model. Different algorithms may better deal with a certain characteristic than others. And the quality of the discovered model may also depend on the predominance of certain characteristics. Performing a comparison without acknowledging the influence of these behavioral characteristics can lead to inconclusive results.
  \item They use manually created processes to generate artificial event data. As a result the studied process characteristics are not randomly included in the processes. Furthermore, relatively few processes and event logs were created. This prevents the results from being statistically and generally valid.
\end{enumerate}
This paper tries to overcome these limitations by proposing a framework that
\begin{enumerate}
  \item abstracts from the modeling notation employed;
  \item starts from the definition of a process population where the probability of several behavioral characteristics can be varying. From this population a random sample of process models and event logs is drawn, thus making it possible to evaluate and generalize the influence of behavioral characteristics on the quality of the discovered models by the different algorithms under analysis;
  \item performs experiments on random samples of a user-specified size, so as to return statistically valid results;
\end{enumerate}

In a nutshell, our framework is based on a classification perspective to evaluate the quality of a discovered model. The framework starts with artificially generating random samples of process models from a specified population of processes. For each model, we generate a training log with fitting traces (to discover a model) and a test log with both fitting and non-fitting traces. Then, the quality of a discovery algorithm with respect to the event log is related to the ability to correctly classify the traces in the test event log: the discovered model should classify a trace representing real process behavior as fitting and a trace representing a behavior not related to the process as non-fitting. In this way the classification approach allows us to evaluate discovery algorithms generating models in different modeling notations because the quality measurement is not based on one specific modeling notation. Furthermore, by using (large) samples of randomly generated models and logs we can make general statements about populations of models and logs.

Obviously, repeating the generation of event logs and process models cannot be done manually to get significant results. We aim at thousands of models and logs in order to generalize. Fortunately, this can be automated through the use of scientific workflows. Scientific Workflow Management (SWFM) systems help users to design, compose, execute, archive, and share workflows that represent some type of analysis or experiment. The advantages of using scientific workflows for process mining are discussed in~\cite{bolt_scientific_2015}.

This framework clearly enables the sensitivity analysis mentioned above: by defining and then sampling from a population of processes using the process-characteristic dimensions, one can evaluate the impact of the different behavioral characteristics on the quality of the models discovered by the different techniques under analysis. This means that the quality measure needs to be returned for each combination of event log and algorithm.

In summary, this paper reports on the formalization and the operationalization of our framework using RapidProM, a scientific workflow system with process mining features. The experiments report the results of their application to four state-of-the-art discovery algorithms. It is beyond the scope of this paper to extensively cover every existing discovery algorithms. However, the operationalization and the experiments show how easy it is to extend to other algorithms. Ultimately, this paper aims to advance the state-of-the-art for evaluating process discovery algorithms.

The remainder of the paper is structured as follows. Section~\ref{sec:contributions} discusses the new evaluation framework including the methodological foundations and all the building blocks. Next, the experiments Section~\ref{sec:validation} describes the validation of the new framework using a large experiment. Section~\ref{sec:discussion} includes a discussion of the experimental results and future research opportunities. Related work is discussed in Section~\ref{sec:related_work} and Section~\ref{sec:conclusions} wraps up the paper with the conclusions.

\section{Discovery Evaluation Framework}
\label{sec:contributions}
This paper reports on a framework that aims to evaluate the quality of discovery algorithms to rediscover a model when confronted with a fraction of its behavior. The framework is designed based on the principles of scientific workflows and experimental design. The former captures the complete evaluation experiment in a workflow that can be automated, reused, refined and shared with other researchers~\cite{barker_scientific_2007}. The latter allows for precise answers that a researcher seeks to answer with the evaluation experiment~\cite{kirk_experimental_1982}.

\begin{figure*}[!t]
\includegraphics[width=1.0\textwidth]{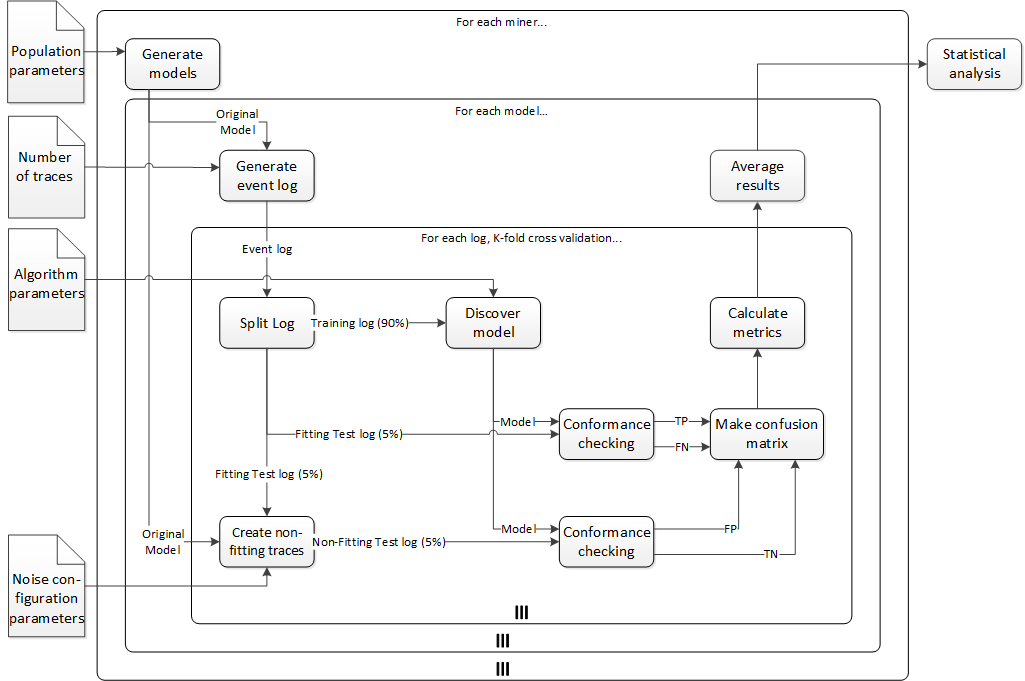}%
\caption{Framework for process discovery algorithm evaluation}%
\label{fig:framework}%
\end{figure*}

To integrate the steps needed for empirically evaluating process discovery algorithms, the framework is built as a scientific workflow. Generally such workflows are represented as a directed graph with nodes denoting computational steps and arcs expressing data flows and data dependencies between steps~\cite{mcphillips_scientific_2009} (see Fig.~\ref{fig:framework}). Bolt et al.~\cite{bolt_scientific_2015} have described generic process mining building blocks to conduct process mining experiments using scientific workflows.

Scientific workflows offer several advantages over traditional ways to conduct process discovery evaluation. The first advantage comes from workflow automation. Experiments evaluating discovery techniques involve large-scale and computationally expensive experiments that require intensive human assistance~\cite{bolt_scientific_2015}. Therefore, automating these experiments removes the need for the human assistance and reduces the time needed to perform experiments. A second benefit comes from the modularity of the workflows. This allows researchers to adapt and extend an existing workflow, e.g. by using other parameter settings or adding new process discovery techniques. A final benefit of scientific workflows is that they can be shared with other researchers. As a result other researchers can replicate experiments with little effort. In this way, our framework facilitates repeated process discovery evaluation, e.g. it becomes trivial to evaluate another set of algorithms or to assess the algorithm's performance with regard to other data characteristics (e.g. noise, control-flow patterns, etc.).

An evaluation analysis aims to test statistical hypotheses about a discovery algorithm. For example, does the presence of loops cause the Alpha+ miner~\cite{Alpha} to discover models with lower fitness? Or: does the Alpha+ miner and Heuristics miner~\cite{Heuristics} perform equally in the fitness dimension on event logs with \textit{non-exclusive choice (OR)} behavior? This makes it fit within the experimental design methodology in which the primary goal is to establish a causal connection between the independent (algorithm, log characteristics) and dependent (model quality criteria) variables~\cite{kirk_experimental_1982}. The three cornerstones of good experimental design are: randomization, replication and blocking~\cite{fisher_design_1960}. The three cornerstones together are fundamental to make the experiments scientifically sound (e.g., avoid bias or wrong conclusions). Therefore, the evaluation framework incorporates each of the cornerstones.

Randomization involves the random assignment of subjects to the treatment in order to limit bias in the outcome of the experiment~\cite{kirk_experimental_1982,voss_design_1999}. In the evaluation context, the subjects are the event logs and the treatments are the discovery algorithms. Therefore, the evaluation framework has to ensure that the event logs generated in the data generation step are random observations from a population of processes with all desired control-flow characteristics. Replication means that more than one experimental unit is observed under the same conditions. It enables researchers to estimate error effects and obtain a more precise estimate of treatment effects~\cite{kirk_experimental_1982}. In the context of process discovery this implies that one needs to test a specific algorithm on more than one event log to accurately assess the effect of that algorithm on model quality. The framework requires that the evaluation is based on a sample of event logs from a given population to obtain better estimates of the studied effect. Finally, blocking an experiment is dividing the observations into similar groups. In this way one can compare the variation between groups more precisely~\cite{voss_design_1999}. For example, if the experiment studies the effect of loops on model quality, also other characteristics such as infrequent behavior could have an effect. Therefore, the evaluation framework allows to vary the presence of loops in models (variable of interest) while holding the infrequent behavior constant to obtain precise estimates of the effect of loops on model quality (studied effect).

The remainder of the section will discuss the design of the framework and its building blocks in more detail.

\subsection{The Design and Use of the Evaluation Framework}
\label{sec:designUse}

The framework focuses on evaluating control-flow discovery algorithms. Therefore, other process related perspectives, such as data and resources, are out of its scope\footnote{However, the same ideas can be applied to include these other perspectives.}. Moreover, the framework aims at evaluation instead of predicting the best performing algorithm given an event log. The framework enables two main objectives: either benchmarking different discovery algorithms, either performing sensitivity analysis, i.e. what effect does a control-flow characteristic or event log characteristic have on algorithm performance.

Fig.~\ref{fig:framework} illustrates the design of the new evaluation framework as a workflow. The directed graph shows how the different tasks needed for evaluating process discovery algorithms are connected. The framework enforces the consecutive execution of data generation, process discovery, quality measurement and statistical analysis. The framework applies a classification approach to allow for the evaluation of discovery algorithms generating models in different notations.

The first step, i.e. the data generation, is triggered by the objective of the experiment. As a result, the objective determines the control-flow behavior a researcher wants to include in the event logs. The specification of control-flow behavior defines a population of process models. This population definition is the start of the data generation phase.

For each discovery algorithm to be tested, multiple instances of the ``generate models'' task run in parallel. The generation results in multiple \emph{random} samples of process models from the same population. Each model (``original model'') is then simulated by the task ``generate event log'' to create one event log, i.e. a \emph{random} sample of traces from all possible traces allowed by the model. The samples of process models and event logs constitute as ``the ground truth''.

Next, the 10-fold cross-validation splits each event log into ten subsets, i.e. folds, of equal sizes. Nine folds form the training log, while the remaining fold serves as the test log. The task ``discover model'' applies the algorithm to induce a model from the training log. To this point, the test log only contains positive examples, i.e. traces that fit the original model. The classification approach requires also negative examples, i.e. traces that do not fit the original model. To generate negative examples, the task ``create non-fitting traces'' alters half of test traces until they cannot be replayed\footnote{Replay uses the trace and the model as input. The trace is ``replayed'' on top of the model to see if there are discrepancies between the trace and the model~\cite{van_der_aalst_process_2016}.} anymore by the original model (``the ground truth'') to create non-fitting traces.

Subsequently, the framework measures the quality of the discovery algorithm by using the discovered model to classify the test traces. This classification happens within the ``conformance checking'' block which replays all traces on the discovered model. A trace representing real process behavior should be classified as allowed, i.e. completely replayable. A trace representing behavior not related to the real process should be classified as disallowed by the discovered model, i.e. not completely replayable. This approach allows for any discovery algorithm generating models with formal replay semantics.

The classification results are then combined in a confusion matrix (see Section~\ref{buildingBlocks}). Based on that matrix, one can compute the well-known recall and precision metrics to evaluate the quality of the discovery algorithm. The framework repeats the process of splitting, discovery, creating non-fitting traces and conformance checking ten times, each time with a different fold as ``test log''. The task ``average results'' computes the average of the metric values over the ten folds to get an estimate of the algorithm's performance. Note that by using 10-fold cross validation the obtained estimate is less likely to suffer from bias, i.e. it helps to decrease the difference of the estimate from the real unknown value of the algorithm's performance on the population of processes. Finally, the task ``statistical analysis'' tests the hypotheses formulated in the context of the objectives.

This framework's design has the property that no two discovery algorithms are applied on the same event log. Furthermore, for each generated model - randomly drawn from a predefined population of models - we randomly draw only a single event log. Consequently, all discovered models and any corresponding quality metric are independent observations which is an important assumption underlying many standard statistical techniques. We acknowledge that this design decision is not the only option as one could test discovery algorithms on the same logs. This alternative design would have more statistical power for the same sample size, however it requires more complex statistical techniques to deal with the dependence between observations. We can compensate for the loss in power in our design by defining the desired power in advance and calculate the sample size required for such power.

Finally, the framework's design based on the experimental design principles enable users to obtain algorithm's performance measures that are \emph{independent} from specific process models and event logs. More specifically, this starts from the generation of (preferably large) random samples of process models and logs from a population, which are then used to estimate the performance of a algorithm with regard to that population. This contrasts evaluation based on a small non-random sample of (manually created) process models and event logs as it could influence the performance estimate to only reflect these particular models and logs.

The following subsection will elaborate on each of the tasks in the evaluation framework.

\subsection{The Building Blocks of the Framework}
\label{buildingBlocks}

\subsubsection{Generate models}
\label{sec:generatemodels}

This building block generates a random sample of process models from a population of models. The inputs of this block are the population characteristics. The user can specify the population by assigning probabilities to each of the model building blocks and setting the size of the models in terms of visible activities. The probabilities of the control-flow characteristics influence the probability for each characteristic to be included in the resulting process model. For example, if the probability of loops is 0.2, then on average 20\% of the model constructs will be of type loop.

In particular, this block allows one to generate models that can feature the basic patterns identified in~\cite{russell_workflow_2006}, namely:

  \paragraph*{-Sequence} Certain process activities need to be sequentially executed.

  \paragraph*{-Exclusive choice} Certain process parts/branches of the process are mutually exclusive. In several notations, this is known as XOR split/join.

  \paragraph*{-Parallelism} Certain parts/branches are ``parallel'', indicating that the activities of a first part of the model do not impose ordering constraints on the activities of a second part. In several notations, this is known as AND split/join.

  \paragraph*{-Inclusive choice} When reaching given points of the process, a choice needs to be made on which part(s) of process that follow need to be carried on. Differently from exclusive choice, multiple parts can be executed in parallel; different from the parallelism construct, not every part that follows the reached point needs to be executed. In several notations, this is known as OR split/join.

  \paragraph*{-Loop} Certain parts of the process can be sequentially repeated multiple times.

This set of pattern is complemented by a number of more advanced patterns:

  \paragraph*{-Silent Transitions} Certain transitions are inserted into the model for a process-routing purpose. For instance, combined with exclusive choices, silent transitions enable certain parts of the process to be skipped.

  \paragraph*{-Duplicate activities} The same activity appears in different parts of the process, indicating that the activity can reoccur.

  \paragraph*{-Long-term dependency} The choice of one or multiple branches at a certain moment in the execution of the process can influence which choices become available at a later point.

  \paragraph*{-Infrequent Paths} This is always combined with an exclusive choice. When the execution reaches an exclusive choice, certain potential process branches have higher chance to be chosen. In fact, this pattern is rather related to the generation of event logs.

These constructs are those typically discovered by discovery algorithms because they are the most relevant. BPMN and other modelling notations support more complex constructs, such as multiple instances and terminating events; however, at the best of our knowledge, no discovery algorithms support their discovery.

As a result, this block allows users to fully control the control-flow behavior in the generated models and generalize the results to the pre-defined population. The user defines a population of process models by setting the following parameters:
\begin{itemize}
  \item Model Size Parameters: mode; min; max.
  \item Control-flow Characteristic Probabilities: sequence; exclusive choice; parallelism; loop; OR; silent transitions; duplicate activities; long-term dependency; infrequent paths
\end{itemize}
In the framework, process models are generated as process trees~\cite{van2011towards}, which support for all the constructs/patterns mentioned above.
To feature the artificial, random generations of process trees, the framework leverages on the technique and implementation reported in\cite{jouck_ptandloggenerator:_2016}, to which interested readers are referred for further information.

\subsubsection{Generate log}\label{generateLogs}

For each generated model the ``generate log'' block creates an event log, i.e. a random sample of all possible traces allowed by that model. This building block simulates the given model to generate a user-specified number of traces per event log. The exclusive choices in each of the models have output-branch probabilities. As a result, the resulting event log contains a random set of fitting and complete traces. The presence of infrequent paths will make some traces more probable than others which will result in event logs with infrequent behavior.

\subsubsection{Split log}

This building blocks applies the first step needed for the 10-fold cross validation evaluation method. The step splits a given event log into ten subsets (folds) of equal size. Nine folds form the ``training log'' and are the input of the discovery algorithm. The tenth fold is the ``test log'' which is split in half: one half constitutes the ``fitting test traces'', the other half will serve as input of the ``create non-fitting traces'' block to make ``non-fitting test traces''. This is repeated ten times such that each of the ten folds becomes a ``test log'' exactly once.

\subsubsection{Create non-fitting traces}
\label{sec:unfittingTraces}

In a classification approach the ``test log'' should contain positive and negative examples. To this point, there are only positive examples, i.e. traces that fit the original model. The "Create non-fitting traces" building block alters the given test traces so that they do not fit the original model anymore. The goal of the non-fitting traces is to punish overgeneralization of discovery algorithms. The flower model is an example of extreme overgeneralization that allows every possible trace involving the set of activities but provides no added value in a business context~\cite{van_der_aalst_process_2016}. Therefore, this paper aims to punish typical overgeneralizing patterns: unnecessary loops, activity skips and parallelism, by altering the traces using specific \textit{noise operations} (see description below) that can add or remove behavior. Additionally, the traces are altered but kept as close to the original trace as possible. In this way, the framework avoids non-fitting traces that would be trivially rejected by underfitting models.

Given a process model and a set of fitting traces, noise is added to each trace as follows. First, one or more of the following noise types based on~\cite{maruster_machine_2003}, is added with a user specified probability:
\begin{itemize}
    \item \textit{Add activity}: one of the process activities is added in a random position within the trace.
    \item \textit{Duplicate an activity}: when an activity is duplicated, it is inserted immediately after the original.
    \item \textit{Remove an activity}: a single activity is randomly removed from the trace.
    \item \textit{Swap consecutive activities}: a random pair of consecutive activities are swapped within the trace.
    \item \textit{Swap random activities}: similar to the previous type of noise, but the activities to be swapped are selected from random positions in the trace (not necessarily consecutive).
\end{itemize}
Then, the modified trace is checked for fitness with respect to the original model.
If the trace does not fit anymore, it is a noisy trace which will not be edited anymore.
If the trace still fits the model, noise is added again (and checked afterwards) until it does not fit anymore, or until noise has been added five times.
If the noisy trace still does not fit the model, the trace is discarded and another trace is randomly selected from the set of fitting traces. This trace follows the same process described above.

\subsubsection{Discover process model}
\label{sec:discoverModel}

This block applies a discovery algorithm to the ``training log'' to induce a process model. This could be any discovery technique with user specified parameter settings. The discovered model will be used for conformance checking.

\subsubsection{Conformance checking}
\label{sec:conformanceChecking}

The conformance checker will replay the given traces on the discovered model. Because the framework applies a classification approach, the replay assigns each trace to a binary class: if a trace can be completely replayed by the discovered model it belongs to the ``fitting'' class, otherwise the trace is part of the ``non-fitting'' class. The number of classes could be extended to create a more fine-grained evaluation. However, we argue that determining the classes for partially fitting traces would require additional research, which is outside the scope of this paper.

\subsubsection{Calculating performance metrics}

The framework summarizes the performance of an algorithm using three standard metrics adopted from the data mining and information retrieval domain: precision, recall and F measure. Traditionally these metrics are based on:

\begin{itemize}
\item True Positives: the number of real traces that \textbf{fit} the discovered model.
\item False Positives: the number of false traces that \textbf{fit} the discovered model.
\item False Negatives: the number of real traces that \textbf{do not fit} the discovered model.
\item True Negatives: the number of false traces that \textbf{do not fit} the discovered model.
\end{itemize}

The \textit{precision} metric refers to the percentage of traces that fit the \textbf{original} model from all the traces that fit the \textbf{discovered} model.
\begin{equation}
\text{Precision} = \frac{\text{True Positives}}{(\text{True Positives} + \text{False Positives})}
\end{equation}

Inversely, the \textit{recall} metric refers to the percentage of traces that fit the \textbf{discovered} model from all the traces that fit the \textbf{original} model.
\begin{equation}
\text{Recall} = \frac{\text{True Positives}}{(\text{True Positives} + \text{False Negatives})}
\end{equation}

The framework uses the \textit{$F_1$} variation of the F measure. This statistic refers to the harmonic average of the precision and recall metrics.
\begin{equation}
\text{$F_1$} = \frac{2 \cdot \text{Precision} \cdot \text{Recall}}{(\text{Precision} + \text{Recall})}
\end{equation}

\subsubsection{Result analysis}

The evaluation framework allows users to compare the performance of algorithms and to study the effect of control-flow characteristics on algorithm performance. The statistical analysis based on the evaluation results depends on the objectives of the experiment and the corresponding hypotheses to be tested. Therefore, the framework does not incorporate specific statistical techniques, instead it can be used with a whole range of exploratory, descriptive and causal statistical techniques to test any hypothesis that can be expressed in terms of precision, recall, $F_1$ score, and characteristics of log and model. The authors believe that this will benefit the adoption of the framework for all types of evaluation studies, rather than serve a specific purpose.

\subsection{Extensibility of the Framework and BPMN}

As claimed in Section 1, the framework reported in this paper is not bound to Petri nets or any other modelling notation. As a consequence, it is extensible to incorporate new discover algorithms, independently of the notations in which these algorithms generate the model. Every change that is necessary to evaluate a new discovery algorithm that produces models in a notation N (say BPMN) is related to the implementation, whereas the framework work-flow does not require changes.

In the implementation, it is necessary to (1) plug-in the new algorithm as a new instantiation of block \emph{Discover model} in Figure~\ref{fig:framework} (cf.\ Section~\ref{sec:discoverModel}) and (2) plug-in a new conformance checker for notation N, with the latter not being necessary if notation N is already among those available in the framework. It is not necessary to change the instantiation of block \emph{Generate models} in Figure~\ref{fig:framework}. Any model generator in any notation that can represent the patterns defined in Section~\ref{sec:generatemodels}, such as process trees, can be employed. These models are only used to generate the event logs with fitting and non-fitting traces and are not directly compared with the models that are discovered.

Consider the case that one wants to evaluate algorithms that discover BPMN models while limiting the number of changes to the current implementation. The implementation of the algorithm needs to be plugged into RapidProM. Also, a conformance checker of BPMN models needs to be available in the implementation. As a matter of fact, this conformance checker is already available in the implementation. First, the BPMN model is converted into a trace equivalent Petri net that is trace equivalent: each execution of the BPMN is possible in the Petri net, and vice versa~\cite{DBLP:conf/bpm/KalenkovaLA14}. Second, the Petri-net conformance checker can be employed. The trace equivalence between the BPMN and the Petri net models guaranteed that every trace that is diagnosed as fitting/unfitting using the equivalent Petri net will also be as such with respect to the original BPMN model.

\section{Implementations and Experiments}
\label{sec:validation}

The framework was operationalized through \emph{RapidProM} extension of the \emph{RapidMiner} analytic-workflow tool~\cite{bolt_scientific_2015}, which contains all the operators mentioned in Section~\ref{buildingBlocks} (see Appendix~B for more details on the implementation).
The experiments were based on Alpha+ Miner~\cite{Alpha}, Heuristics Miner~\cite{Heuristics}, ILP Miner~\cite{ILP} and Inductive Miner~\cite{InductiveMiner}.
These discovery algorithms are returning Petri nets, which require a suitable conformance checker. The choice has fallen on the alignment-based \emph{conformance-checking} technique presented in~\cite{Mannhardt2016}, which is available in RapidProM and, differently e.g.\ from the token-based algorithm~\cite{rozinat_conformance_2008}, is able to deal with invisible transitions and duplicate activity labels.
Excluding the ILP miner, the other algorithms were used with the default configuration.
The ILP miner was configuration to generate models in which the final marking is the empty marking (no tokens remaining). Any other configuration generates process models in which the ILP miner does not state what the final marking is, which would require a model inspection by a human. The human involvement would hinder the possibility of an automatic workflow.

The generations of process models and event logs are based on the techniques and implementations that are respectively reported in~\cite{jouck_ptandloggenerator:_2016} and in~\cite{jouck_simulating_2017}, which are in line with Section~\ref{buildingBlocks}.

We conducted two rounds of experiments. The first round validates the usefulness of the proposed framework through an experiment consisting of a detailed empirical analysis of the process discovery algorithms mentioned above. The discussion of the first round is in Section~\ref{ExperimentalSetup} and the experimental results are reported in Section~\ref{AnalysisOfResults}. In the second experiment round, the flexibility of the framework and its support for large-scale experiments is validated by extending the first round to experiments five times larger. Section~\ref{sec:ExperimentConclusions} reports on the second round.

\subsection{Setup of the First Experiment}
\label{ExperimentalSetup}

As mentioned in Section~\ref{sec:designUse}, the goal of this framework is to analyze and compare the accuracy of process discovery techniques to rediscover process models based on observed executions (i.e., event logs).
The population of process models that we aim to rediscover is generated by varying a number of parameters, which identifies the probability of occurrences of typical process characteristics, such as parallel branches, silent transitions and infrequent paths. Section~\ref{buildingBlocks} has discussed the constructs which, so far, our framework allows for and how the probabilities influence the generated process models.
In the first round of experiments, the population of models is generated by varying the probability of duplicate activities and by enabling or disabling the presence of infrequent paths. In this way, we can study the impact of infrequent behavior and of different frequencies of duplicate activities on the accuracy of process discovery techniques. Section~\ref{sec:ExperimentConclusions} will report on the extended experiment where the probability of the other process characteristics are also varied.

Therefore, the experimental design includes all the combinations of three independent variables: process discovery technique used, presence or absence of infrequent behavior and the probability of having \textit{duplicate activities}. The three variables and their levels are summarized in Table~\ref{tab:experimentSetup}. In total, the 56 possible combinations are included in the experiment: 4 discovery techniques $\times$ 2 levels of infrequent behavior $\times$ 7 probabilities of duplicate activities.

As mentioned above, the other process characteristics are not taken into account in this analysis. The probability of non-exclusive choice (OR) and of loops
are set to zero and, hence, these two constructs do not occur. The probability of sequence, exclusive choice and parallelism is set and kept fixed to values 46\%, 35\% and 19\%, respectively. These values have been determined after analysing their frequencies in the large collections of models reported in~\cite{kunze_towards_2011}. In this work, Kunze et al.\ have observed that 95\% of the models consist of activities connected in sequences, 70\% of the models consist of activities, sequences and XOR connectors and 38\% consist of sequences, activities and AND connectors (see Fig.~4b of the paper). Assuming independence of occurrence probability of sequences, AND and XOR, it follows that:
\begin{equation*}
\begin{small}
\begin{array}{l}
  P(sequence)=0.95 \\
  P(sequence \land XOR)=P(sequence) \times P(XOR) = 0.70 \\
  \quad \Rightarrow P(XOR)=0.74 \\
  P(sequence \land AND)=P(sequence) \times P(AND) = 0.38 \\
  \quad \Rightarrow P(AND)=0.4 \\
\end{array}
\end{small}
\end{equation*}
When these values are normalized to 1, the final probabilities of the constructs are obtained.

For each discovery technique a random sample of 62 process models is drawn. The sample size of 62 models allows us to study the effect of process discovery techniques, infrequent paths and different probabilities of duplicate-activity occurrences (and their interactions) using a fixed effects ANOVA analysis~\cite{voss_design_1999} with significance level $\alpha = 0.05$ and power $1-\beta = 0.98$\footnote{The power was computed with the G*Power tool~\cite{faul_statistical_2009}}. This power indicates the probability to detect a significant effect when two mining algorithms actually differ by a relatively small difference. In total (i.e., sum of all combinations), 3472 process models were generated.

\begin{figure}[t]
  \centering
  \includegraphics[width=\columnwidth]{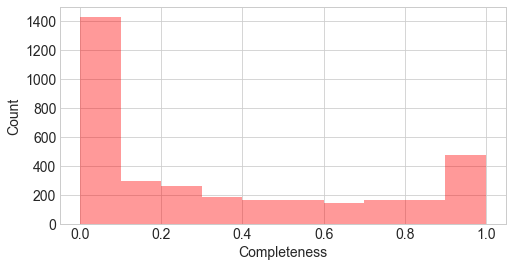}
  \caption{Distribution of completeness of logs wrt.\ their respective process models. Completeness is measured as the fraction of traces allowed by the model that are present in the event log.}
  \label{fig:histogram}
\end{figure}

For each of the obtained process models, an event log containing between 200 and 1000 traces is generated (See Section~\ref{generateLogs}). For each generated log, we can calculate the completeness, i.e. the ratio of unique traces in the log to all possible unique traces according to the model using the technique described in~\cite{janssenswillen_calculating_2016}. Fig.~\ref{fig:histogram} shows that the completeness varies between 0 and the maximum of 1. The experiment applies the following discovery techniques with its default parameter settings: Alpha+~\cite{Alpha}, Heuristics~\cite{Heuristics}, ILP~\cite{ILP} and Inductive Miner~\cite{InductiveMiner}. As such, the choice of techniques covers most families of discovery approaches {}(see~\cite{van_der_aalst_process_2016}).

\begin{table}[t]
  \caption{Summary of the Possible Values of the Four Variables Included in the Experimental Setup: 56 ($4 \times 2 \times 7$) value combinations. The probability of Duplicate Activities indicates the average percentage of duplicated visible activity labels in the process model.}
  \label{tab:experimentSetup}
  \centering
  {\renewcommand{\tabcolsep}{0.3cm}
    \begin{tabular}{cccc}
      \toprule
      \textbf{Discovery} & \textbf{Infrequent} & \textbf{Probability}\\
      \textbf{Technique} & \textbf{Paths} & \textbf{Duplicate Activities}\\
      \midrule
      Alpha+~\cite{Alpha} & False & 0.0, 0.05 \\
      Heuristics~\cite{Heuristics} & True & 0.10, 0.15 \\
      ILP~\cite{ILP} & & 0.15, 0.20 \\
      Inductive~\cite{InductiveMiner} & & 0.25, 0.30 \\
      \bottomrule
    \end{tabular}}
\end{table}

\subsection{Analysis of the Results of the First Experiment}
\label{AnalysisOfResults}
This section reports on the results of the experiments illustrated in Section~\ref{ExperimentalSetup}.

The effect of process-discovery techniques, infrequent paths and different probabilities of duplicate-activity occurrences can be analyzed using one-way ANOVA analysis if the assumptions of homogeneity of variances and normality of the dependent variable hold~\cite{voss_design_1999}. However, both assumptions were violated for every dependent variable, i.e. $F_1$, recall and precision. Therefore, the non-parametric \textit{Kruskall-Wallis} test (KW)~\cite{siegel_nonparametric_1988} was applied instead.

KW is used for testing whether $k$ independent samples are from different populations. It starts by ranking all the data from the different samples together: assign the highest score a rank 1 and the lowest a rank $N$, where $N$ is the total number of observations in the $k$ samples. Then, the average ranking for each sample is computed, e.g. the mean of sample $j$ is denoted as $\bar{R}_j$. With $n$ the number of observations in each sample, the test statistic KW, which follows a $\chi^2$ distribution with $k-1$ degrees of freedom, can be calculated as~\cite{siegel_nonparametric_1988}:

\begin{equation}
\text{KW} = [\frac{12}{(N(N+1))}\sum_{j=1}^{k}{n\bar{R}_j^2}]-3(N+1)
\end{equation}

If the calculated KW is significant, then it indicates that at least one of the samples is different from at least one of the others. Subsequently, the multiple comparison post hoc test is applied to determine which samples are different. More specifically, for all a pairs of samples $R_i$ and $R_j$ it is tested whether they differ significantly from each other using the inequality~\cite{siegel_nonparametric_1988}:

\begin{equation}
|R_i - R_j| \geq z_{\alpha/k(k-1)}\sqrt{\frac{N(N+1)}{12}(\frac{2}{n})}
\end{equation}

The $z_{\alpha/k(k-1)}$ value can be obtained from a normal distribution table given a significance level $\alpha$. The formula adjusts this $\alpha$ with a Bonferroni correction to compensate for multiple comparisons. If the the absolute value of the difference in average ranks is greater than or equal to the critical value, i.e. the right side of the equation, then the difference is significant.

Finally, the Jonckheere test~\cite{siegel_nonparametric_1988} can be used to test for a significant trend between the $k$ samples. First, arrange the samples according to the hypothesized trend, e.g. in case of a positive trend from smallest hypothesized mean to highest hypothesized mean. Then count the number of times an observation in sample $i$ precedes an observation in sample $j$, denoted as $U_{ij}\ \forall i<j$. The Jonckheere test statistic $J$ is the total number of these counts:

\begin{equation}
\text{J} = \sum_{i<j}^{k}{U_{ij}}
\end{equation}

When $J$ is greater than the critical value (see~\cite{siegel_nonparametric_1988} for the sampling distribution) for a given significance level $\alpha$, then the trend between the $k$ samples is significant.

\subsubsection{The Effect of Process Discovery Technique}

The goal is to learn the effect of a process discovery technique on each of the dependent variables: recall, precision and $F_1$ score. The other variables (i.e., infrequent paths level and probability of duplicate activities) are part of the error term.

We apply the KW method, to test whether the average rank differs between the four process discovery techniques (i.e. samples). In this case we ranked all the 3472 averages over the 10-fold cross validation for recall, precision and $F_1$ values ignoring sample membership (i.e. discovery technique). The highest value for recall, precision and $F_1$ gets rank 1 (lowest rank), while the lowest absolute value gets rank 3472 (highest rank). Then we computed the average ranking per miner, i.e. the average position of a discovered model by that miner for that quality metric on a scale from 1 to 3472. A higher average ranking means worse performance. The ranking summary is shown in Table~\ref{tab:AverageRanksPerMiner}.

\begin{table}
  \caption{Average Ranks per Miner. Each cell indicates the average ranking for a specific performance dimension (row header) and for a specific miner (column header). One can compare miners by comparing the average ranks within one row.}
  \label{tab:AverageRanksPerMiner}
  \centering
  {\renewcommand{\tabcolsep}{0.2cm}
    \begin{tabular}{lcccc}
      \toprule
      & \textbf{Alpha+} & \textbf{Heuristics} & \textbf{ILP} & \textbf{Inductive} \\
      \midrule
      \textbf{Recall} & 2361.94 & 2650.35 & 505.99 & 1427.73 \\
      \textbf{Precision} & 2155.57 & 2624.42 & 1007.66 & 1158.35 \\
      \textbf{$F_1$ score} & 2318.14 & 2646.44 & 697.00 & 1284.42 \\
      \bottomrule
    \end{tabular}}
\end{table}

Based on the average rankings in Table~\ref{tab:AverageRanksPerMiner}, the order suggested between process discovery techniques is: ILP $>$ Inductive $>$ Alpha+ $>$ Heuristics for recall, precision and $F_1$ scores. It means that the ILP miner creates the best models in terms of recall, precision and $F_1$ scores (see Section~\ref{sec:discussion} for an elaborate discussion). The Inductive miner outperforms the Alpha+ miner, which in turn outperforms the Heuristics miner. The results of the KW test confirm that the differences in average rankings between the four miners are statistically significant (significance level $\alpha = 0.05$). Moreover, the multiple comparison post-hoc test (cf. supra) also confirms the statistical significance of the differences between algorithms. See Table~1 in Appendix~A for a summary of the statistical test results for the $F_1$ scores.

\subsubsection{The Effect of Infrequent Paths}

The analysis tests whether the presence/absence of infrequent paths\footnote{Infrequent paths are denoted with an imbalance in execution probabilities of the output-branches of each exclusive choice construct in the model which results in an event log containing infrequent behavior.} has an impact on the average ranking of the four process discovery techniques for recall, precision and $F_1$ scores. The effect of duplicate activities is part of the error term.

Firstly, the sample is split into two subsets: experiments with infrequent behavior and experiments without infrequent behavior. This division is called \textit{blocking} (see Section~\ref{sec:contributions}) which is done to isolate the variation in recall, precision and $F_1$ scores attributable to the absence/presence of infrequent paths. Secondly, the KW test is applied to each subset.

Table~\ref{tab:AverageRanksPerMiner2} contains the average rankings per process discovery technique grouped by metric and experiments with and without infrequent behavior. These rankings suggest the same order between process discovery techniques in all cases: ILP $>$ Inductive $>$ Alpha+ $>$ Heuristics. This leads to the assumption that the process discovery techniques are not influenced by the absence or presence of infrequent behavior. Based on the KW and multiple comparison post-hoc test, only the difference between the ILP and Inductive miner in case of infrequent behavior is not statistically significant for precision (see Table~2 in Appendix~A). Therefore, one cannot accept the assumption that infrequent paths do not influence process discovery techniques.

\begin{table}[!t]
    \caption{Average Ranks per Miner per Probability}\label{tab:AverageRanksPerMiner2}
    \begin{subtable}{\columnwidth}
      \caption{Average Ranks per Miner Without Infrequent Behavior}
      \label{tab:WithoutInfrequent}
      \centering
      {\renewcommand{\tabcolsep}{0.2cm}
      \begin{tabular}{lcccc}
        \toprule
         & \textbf{Alpha+} & \textbf{Heuristics} & \textbf{ILP} & \textbf{Inductive} \\
        \midrule
        \textbf{Recall} & 1196.47 & 1338.58 & 262.01 & 676.93 \\
        \textbf{Precision} & 1063.51 & 1317.79 & 485.44 & 607.26 \\
        \textbf{$F_1$ score} & 1180.24 & 1338.25 & 325.71 & 629.80 \\
        \bottomrule
      \end{tabular}}
    \end{subtable}
    \newline
    \vspace*{1 cm}
    \newline
    \begin{subtable}{\columnwidth}
      \caption{Average Ranks per Miner With Infrequent Behavior}
      \label{tab:WithInfrequent}
      \centering
      {\renewcommand{\tabcolsep}{0.2cm}
      \begin{tabular}{lcccc}
        \toprule
         & \textbf{Alpha+} & \textbf{Heuristics} & \textbf{ILP} & \textbf{Inductive} \\
        \midrule
        \textbf{Recall} & 1162.64 & 1313.58 & 242.94 & 754.85 \\
        \textbf{Precision} & 1088.57 & 1306.34 & 523.05 & 556.04 \\
        \textbf{$F_1$ score} & 1136.22 & 1310.40 & 367.71 & 659.93 \\
        \bottomrule
      \end{tabular}}
    \end{subtable}
\end{table}

\subsubsection{The Effect of Duplicate Activities}

The analysis investigates how the accuracy of each process discovery technique (in terms of precision, recall and $F_1$ score) is influenced by the probability of duplicate activities (i.e. the average percentage of duplicated visible activity labels in the process models). The effect of infrequent behavior is part of the error term.

\begin{figure*}[t!]
  \centering
    \begin{subfigure}[b]{1\columnwidth}
      \centering
      \includegraphics[width=1\textwidth]{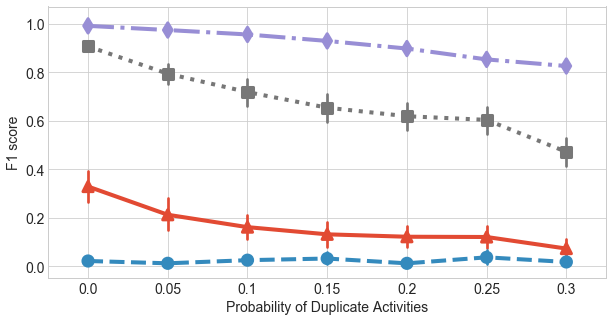}
      \caption{Duplicated Activities}
      \label{fig:ReocurringTask}
      \end{subfigure}
    \begin{subfigure}[b]{1\columnwidth}
      \centering
      \includegraphics[width=1\textwidth]{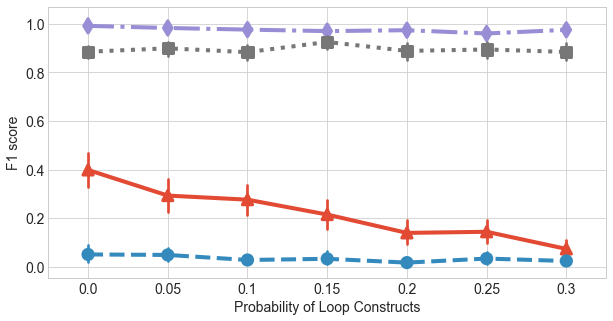}
      \caption{Loops}
      \label{fig:Loop}
    \end{subfigure}
    \hfil
    \begin{subfigure}[b]{1\columnwidth}
      \centering
      \includegraphics[width=1\textwidth]{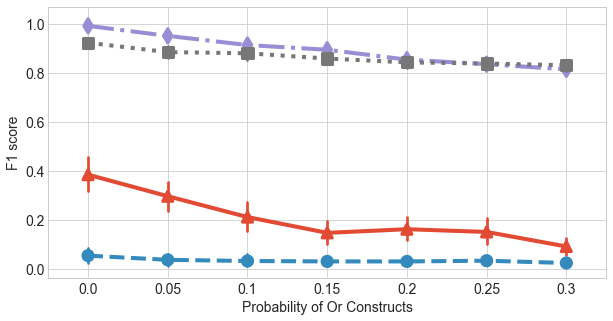}
      \caption{OR}
      \label{fig:InclusiveOR}
    \end{subfigure}
    \begin{subfigure}[b]{1\columnwidth}
      \centering
      \includegraphics[width=1\textwidth]{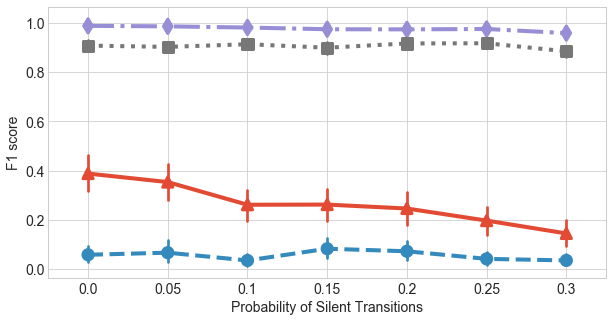}
      \caption{Silent Transitions}
      \label{fig:SilentTransition}
    \end{subfigure}
    \hfil
    \begin{subfigure}[b]{1\textwidth}
      \centering
      \includegraphics[width=0.7\textwidth]{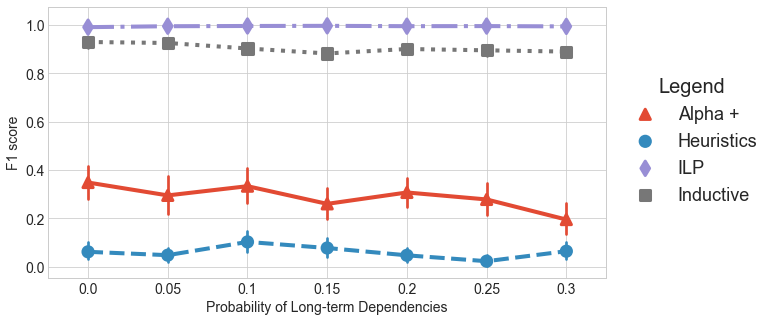}
      \caption{Long-term Dependencies}
      \label{fig:LongtermDependency}
    \end{subfigure}

\caption{$F_1$ scores for Process Discovery Techniques for different probabilities of Process Control-flow Characteristics}
\label{fig:Charts}
\end{figure*}

Fig.~\ref{fig:ReocurringTask} illustrates the average $F_1$ scores for all the process discovery techniques over different probabilities of duplicate activities. This graph indicates a negative trend, i.e. the probability of duplicate activities has a negative effect on $F_1$ scores. To determine whether such a trend is statistically significant, an in-depth analysis is performed.

First, the sample is divided into subsets grouped by process discovery technique. As such, the variation in accuracy associated with the discovery technique is isolated. Then, similar to the analysis above, the KW test is applied to compare the average rankings of the discovered models.

Table~\ref{tab:AverageRanksPerMiner3} contains one subtable for each process discovery technique with the average ranks for all three metrics by the probability of duplicate activities.

\begin{table}[!t]
    \caption{Average Ranks of Process Discovery Techniques per Probability Duplicate Activities}\label{tab:AverageRanksPerMiner3}
    \begin{subtable}{\columnwidth}
        \caption{Average Ranks Alpha+ Miner per Probability Duplicate Activities}
        \label{tab:AverageAlpha}
        \centering
        {\renewcommand{\tabcolsep}{0.1cm}
        \begin{tabular}{lccccccc}
          \toprule
          \textbf{Prob. Duplicate} & \multirow{2}{*}{0} & \multirow{2}{*}{0.05} & \multirow{2}{*}{0.10} & \multirow{2}{*}{0.15} & \multirow{2}{*}{0.20} & \multirow{2}{*}{0.25} & \multirow{2}{*}{0.30} \\
          \textbf{Activities} & & & & & & &\\
          \midrule
          \textbf{Recall} & 339.47 & 417.21 & 428.08 & 451.02 & 435.54 & 479.52 & 490.66\\
          \textbf{Precision} & 346.37 & 421.59 & 423.21 & 449.60 & 431.10 & 478.01 & 491.62\\
          \textbf{$F_1$ score} & 339.46 & 417.56 & 427.86 & 450.78 & 435.69 & 479.36 & 490.78\\
          \bottomrule
        \end{tabular}}
    \end{subtable}%
    \newline
    \vspace*{1 cm}
    \newline
    \begin{subtable}{\columnwidth}
        \caption{Average Ranks Heuristics Miner per Probability Duplicate Activities}
        \label{tab:AverageHeuristics}
        \centering
        {\renewcommand{\tabcolsep}{0.1cm}
        \begin{tabular}{lccccccc}
          \toprule
          \textbf{Prob. Duplicate} & \multirow{2}{*}{0} & \multirow{2}{*}{0.05} & \multirow{2}{*}{0.10} & \multirow{2}{*}{0.15} & \multirow{2}{*}{0.20} & \multirow{2}{*}{0.25} & \multirow{2}{*}{0.30} \\
          \textbf{Activities} & & & & & & &\\
          \midrule
          \textbf{Recall} & 432.08 & 429.27 & 435.26 & 428.28 & 442.85 & 437.97 & 435.79\\
          \textbf{Precision} & 431.96 & 429.40 & 434.90 & 428.14 & 442.76 & 438.51 & 435.83\\
          \textbf{$F_1$ score} & 432.04 & 429.34 & 435.19 & 428.27 & 442.85 & 438.06 & 435.75\\
          \bottomrule
        \end{tabular}}
    \end{subtable}
    \newline
    \vspace*{1 cm}
    \newline
    \begin{subtable}{\columnwidth}
        \caption{Average Ranks ILP Miner per Probability Duplicate Activities}
        \label{tab:AverageILP}
        \centering
        {\renewcommand{\tabcolsep}{0.1cm}
        \begin{tabular}{lccccccc}
          \toprule
          \textbf{Prob. Duplicate} & \multirow{2}{*}{0} & \multirow{2}{*}{0.05} & \multirow{2}{*}{0.10} & \multirow{2}{*}{0.15} & \multirow{2}{*}{0.20} & \multirow{2}{*}{0.25} & \multirow{2}{*}{0.30} \\
          \textbf{Activities} & & & & & & &\\
          \midrule
          \textbf{Recall} & 466.09 & 438.65 & 457.05 & 462.85 & 416.74 & 418.97 & 381.15\\
          \textbf{Precision} & 172.53 & 295.21 & 369.48 & 431.76 & 521.84 & 604.08 & 646.60\\
          \textbf{$F_1$ score} & 185.64 & 288.92 & 370.51 & 430.34 & 519.42 & 602.04 & 644.64\\
          \bottomrule
        \end{tabular}}
    \end{subtable}
    \newline
    \vspace*{1 cm}
    \newline
    \begin{subtable}{\columnwidth}
        \caption{Average Ranks Inductive Miner per Probability Duplicate Activities}
        \label{tab:AverageInductive}
        \centering
        {\renewcommand{\tabcolsep}{0.1cm}
        \begin{tabular}{lccccccc}
          \toprule
          \textbf{Prob. Duplicate} & \multirow{2}{*}{0} & \multirow{2}{*}{0.05} & \multirow{2}{*}{0.10} & \multirow{2}{*}{0.15} & \multirow{2}{*}{0.20} & \multirow{2}{*}{0.25} & \multirow{2}{*}{0.30} \\
          \textbf{Activities} & & & & & & &\\
          \midrule
          \textbf{Recall} & 239.90 & 339.92 & 415.75 & 462.62 & 492.83 & 515.94 & 574.54\\
          \textbf{Precision} & 241.90 & 331.42 & 405.01 & 478.28 & 496.85 & 490.82 & 597.22\\
          \textbf{$F_1$ score} & 219.71 & 339.45 & 410.69 & 474.73 & 500.88 & 505.77 & 590.27\\
          \bottomrule
        \end{tabular}}
    \end{subtable}
\end{table}

For the Alpha+ Miner, the data (shown in Table~\ref{tab:AverageAlpha}) seems to suggest that as the probability of duplicate activities increases, the models generated by Alpha+ miner deteriorate in terms of recall, precision and $F_1$ score. To test this impression statistically, we will rely on the KW and Jonckheere tests. Both tests show that there is statistically significant negative trend in the relative quality of the generated models as the probability of duplicate activities increases. A pairwise comparison of each probability of duplicate activities does not provide a clear picture how this trend looks like for recall, with many comparisons statistically insignificant. For precision and $F_1$ on the other hand, the quality of the models decreases significantly whenever the probability of duplicate activities increases from 0\% to more than or equal to 15\% (see Table~3 in Appendix~A).

The models discovered using the Heuristics Miner seem insensitive to the probability of duplicate activities (see Table~\ref{tab:AverageHeuristics}). The KW and Jonckheere tests confirm that there is indeed statistically insufficient evidence of a trend in recall, precision and $F_1$ score as the probability of duplicate activities increases (see Table~4 in Appendix~A). A possible explanation will be discussed in Section~\ref{sec:discussion}.

The results for the ILP Miner in Table~\ref{tab:AverageILP} suggest a positive trend in the probability of duplicate activities in terms of recall! However, in terms of precision, the ILP miner shows high sensitivity to the probability of duplicate activities. The KW and Jonckheere tests confirm both statements. The pairwise comparisons of duplicate activities reveals the significant negative trend in terms of precision and $F_1$ scores of the generated models as the probability of duplicate activities increases (see Table~5 in Appendix~A).

The findings for the Inductive Miner indicate that as the probability of duplicate activities increases, the model quality in terms of recall, precision and $F_1$ score deteriorates. This effect, though, seems to level off as we reach higher probabilities of duplicate activities. The KW and Jonckheere tests show that there is indeed a significant negative trend in the relative quality of the generated models as the probability of duplicate activities increases. However, at a probability of around 15\% of duplicate activities, this effect seems to have reached a plateau and stays stable (see Table~6 in Appendix~A).

\subsection{Extended Experiment}\label{sec:ExperimentConclusions}

The above experiments have validated the usefulness of the proposed evaluation framework to support the benchmark and sensitivity analysis evaluation objectives. The proposed framework is also flexible as it allows users to easily setup extended experiments. Here, we have extended the above experiment with other control-flow characteristics. The probability of the basic characteristics, sequence, parallel and exclusive choice, is set the same as in the previous experiments. In this experiment, for each process characteristic, we have varied the probability of its occurrence while setting the probability of the others to zero. Instead of 3472 observations as in the first experiment, the extended experiment results in 17360 observations. An in-depth discussion of the results as done for the duplicate activities is not included in this paper due to lack of space. The graphs in Fig.~\ref{fig:Charts} show the average $F_1$ score for all the discovery techniques over different probabilities of inclusion of the control-flow characteristics. Section~\ref{sec:discussion} details a thorough discussion of the results of this extended experiment, along with the first experiment.

\section{Discussion And Future Research}
\label{sec:discussion}
This section discusses the overall results found in the experiments section and describes future research opportunities.
Fig.~\ref{fig:Charts} illustrates how the different algorithms score in terms of $F_1$ score with varying probability of the constructs.

\begin{figure*}[t!]
  \centering
  \includegraphics[width=1\textwidth]{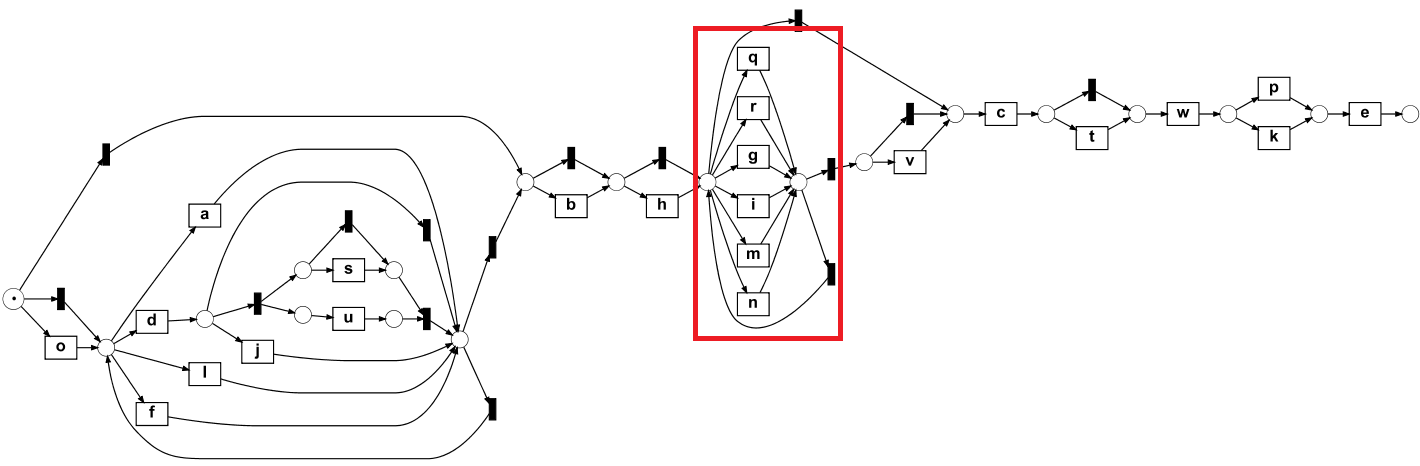}
  \caption{Model Discovered by the Inductive Miner}
  \label{fig:inductive}
\end{figure*}

\begin{figure}[t!]
  \centering
  \includegraphics[width=0.75\columnwidth]{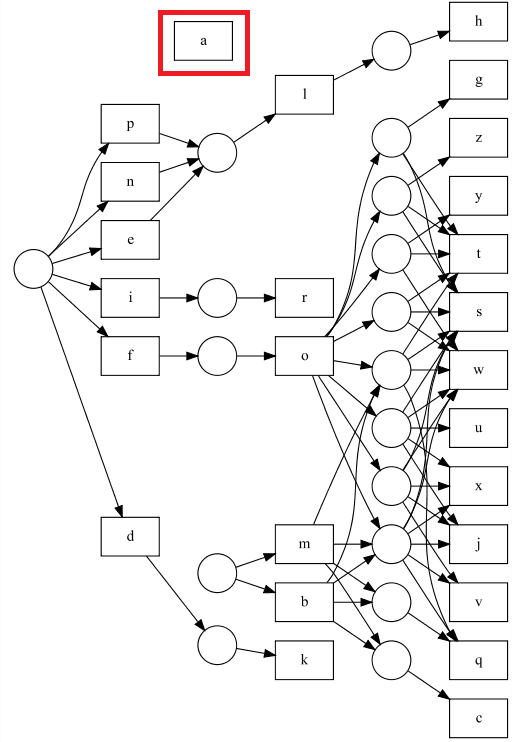}
  \caption{Model Discovered by the ILP miner}
  \label{fig:ilp}
\end{figure}

First, the graphs clearly highlight that ILP and Inductive Miner perform significantly better than Alpha+ and Heuristic Miner. In fact, this is not surprising because the latter two miners are not guaranteed to produce sound models, which allow executions to be carried out till completion. Models discovered with Alpha+ and Heuristic Miner can contain deadlocks, livelocks, and other anomalies~\cite{van_der_aalst_process_2016}. When a model is indeed  not sound, it cannot replay  traces until the end and, hence, the confusion matrix may contain few true positives (often none), causing precision, recall and, hence, $F_1$ scores to be very low (often zero).
This is not trivial because, although the theory already postulated it, it was not clear how much the lack of soundness guarantee was practically affecting the results.
Ultimately this means that Alpha+ Miner and Heuristic Miner can be useful to gain an initial insight into the general structure of the process but cannot be used for more mature answers or for automatically generating models that can be plugged into a Process Management System to enforce certain process behavior.

Looking at Fig.~\ref{fig:Charts}, ILP miner tends to perform better than Inductive Miner in terms of $F_1$ score. This is observed for all constructs and all occurrence probabilities. In particular, for such constructs as silent transitions and long-term dependencies, the $F_1$ score is steadily around 1, which indicates almost perfect precision and recall. This result is far from being trivial: as discussed~\cite{ILP}, the ILP miner focuses on producing models that can replay every trace of the event log, without trying to maximize precision. Furthermore, because ILP miner only aims at replaying the traces in the event log used for discovery, one would expect that a different event log, used e.g.\ for testing, would not let the discovered models score high in recall, either.

The superiority of ILP miner is further supported by visually comparing the models that ILP generates and those from the Inductive Miner, such as the models
in Figures~\ref{fig:inductive} and \ref{fig:ilp} respectively discovered through the Inductive and ILP Miner. The red boxes in the figure illustrate the unprecise parts of the model. For the Inductive-Miner model, the transitions in the box can be executed in any order and, because of the loop, an arbitrary number of time. Of course, in the reality, these transitions should occur in a more precise order; but the miner is unable to ``see it''.
Conversely, for the model discovered through the ILP miner, the only ``source of imprecision'' is related to the ``floating transition'' $a$ but it is just one out of 26 transitions. This does not affect the precision. As discussed in Section~\ref{sec:unfittingTraces}, to punish for imprecise behavior, our framework injects noise into fitting traces. In case of the mode by the ILP miner, the probability that the noise would involve the only ``floating transition'' $a$ is low. On the other hand, the probability that noise affects activities present in precise regions of the model is high. Such deviations in very precise regions are easily detected, resulting in high $F_1$ scores for the ILP miner. The same reasoning is shared among most of models.

Another interesting result for both Inductive and ILP miner is that the values of $F_1$ score seem not to be really affected by the amount of occurrences of the process constructs, except for duplicate activities and, limitedly, from the OR construct. The OR is known to be a hard construct and neither of the two miners provides specific support for it (for Inductive Miner, at least for the version being evaluated). For duplicate activities, this can be explained by the fact that both ILP and Inductive Miner do not natively support mining models where different transitions share the same activity label.
This means that duplicate activities are ``emulated'' through loops and floating transitions (see above),
which would underfit the behavior observed in the event log, thereby yielding low precision.

We acknowledge in this paper that the results are also affected by the fact that training event logs do not contain noise, namely traces that are not generated by the original, artificial models.
As an example, ILP miner tends to be very sensitive to noise: since it discovers models that are able to replay every event trace, if the logs contain noise, the discover models would incorporate behavior that should not be allowed, thus negatively affecting precision. Conversely, Inductive Miner would likely be less affected because it features
some noise detection, able to detect whether a trace is really part of the process or a noise/outlier.
This is based on the frequencies of occurrences of certain patterns in the traces of the event log~\cite{InductiveMiner}.
As future work, we aim to add new ingredients to our analysis and consider a variable percentage of training-log noise and to study how discovery algorithms are affected by the amount of noise, in terms of $F_1$ score.

A future interesting extension to our framework is parameter sensitivity. Every miner that we employed in our experiments can be customized by setting the values of certain parameters. In this paper, we ruled out the parameter sensitivity by using the default parameter values.
For instance, Inductive Miner can be customized by varying the threshold of noise detection, also known in the algorithm as $\alpha$-value, which can vary from zero to one. The model in Fig.~\ref{fig:inductive} was mined with the default $\alpha$-value, which is 0.2, leading to a $F_1$ score of 0.25. For the specific case, we manually reduced $\alpha$ to 0, thus not supporting noise detection. This led to an increase of $F_1$ till a clearly better $0.67$, which is  due to the fact that, as mentioned, the training logs do not contain noise.

However, we believe that not addressing these aspects do not invalidate the arguments stated in this paper. While it is easy to accommodate them in the framework, the current experiments can illustrate that our analysis framework does already properly address the research objectives stated in Section~\ref{sec:introduction}. First, even though we always translate the discovered models to Petri nets, it was related to the fact that the available conformance checker requires it. Our framework is modelling notation independent and as such there is nothing that is specific for Petri nets. Second, the results are based on a statistically-significant sample of models and event logs from a population with several characteristics, which make the results generally valid.

\section{Related work}
\label{sec:related_work}

Several frameworks for evaluating process discovery algorithms have been proposed. Rozinat et al.~\cite{rozinat_towards_2007} introduced the first evaluation framework, Wang et al.~\cite{wang_efficient_2013} and Ribeiro et al.~\cite{ribeiro_recommender_2014,joel_ribeiro_method_2015} extended the Rozinat framework to evaluate and predict the best algorithm. Weber et al.~\cite{weber_framework_2013} proposed an alternative framework that takes a probabilistic perspective. In addition to the evaluation frameworks, De Weerdt et al.~\cite{de_weerdt_multi-dimensional_2012}, Vanden Broucke et al.~\cite{vanden_broucke_uncovering_2014,vanden_broucke_fodina:_2017} and Augusto et al.~\cite{augusto_automated_2017} performed benchmarking studies of process discovery techniques.

As indicated in Section 1, our framework evaluates the quality of models on the basis of measures of precision and recall that are not bound to any modelling notation. Conversely, the existing body of research is based on metrics that are applicable to one notation, mostly Petri net~\cite{bolt_scientific_2015,rozinat_towards_2007,wang_efficient_2013,ribeiro_recommender_2014,joel_ribeiro_method_2015,weber_framework_2013,de_weerdt_multi-dimensional_2012,vanden_broucke_uncovering_2014,vanden_broucke_fodina:_2017,augusto_automated_2017}.
This means that, not only would one need to introduce a new conformance checker for the new notation, but also one would need to redefine the precision and recall metrics. The introduction of a new conformance checker is also necessary in our framework and it cannot be prevented because it is necessary that the checker is aware of the replaying semantics of the notation.

Section 1 also indicated that the second and third advantage of our framework is that it is based on the generation of a sufficiently-large number of artificial models to guarantee a statistical validity of the analysis. Conversely, the existing frameworks base their conclusions on samples that are small, either a few real-life event logs~\cite{de_weerdt_multi-dimensional_2012,ribeiro_recommender_2014,augusto_automated_2017}, either artificial but not randomly generated~\cite{de_weerdt_multi-dimensional_2012,rozinat_towards_2007,wang_efficient_2013,ribeiro_recommender_2014,joel_ribeiro_method_2015,weber_framework_2013,vanden_broucke_uncovering_2014}, thus limiting the statistical validity of the analysis. Also, the artificial process models are not generated by controlling the probability of certain constructs to be present. This means that the event logs generated from these models do not allow one to evaluate the correlation between the quality of the discovered models and the presence of certain process constructs.

Furthermore, all frameworks, except Weber et al.~\cite{weber_framework_2013}, leverage on the typical process-mining notions of precision, generalization and fitness from literature to evaluate the quality of the discovered models (see, e.g.,\cite{van_der_aalst_process_2016}). There is a clear correlation between the precision and recall that we employ and the typical process-mining measures of model quality. However, the process-mining measures are designed considering that the real process model is not known and that one only observes the positive cases, namely the traces that are part of the real process. The negative cases, i.e.\ the executions/traces that do not fit the real process, are not known because they would require to know the real model. Therefore the process-mining measures of model quality try to artificially generate the negative cases based on estimation and, hence, the measure results are estimates (see, in this respect, also~\cite{vanden_broucke_determining_2014}). Since we know the real process model, we can generate both positive and negative traces and label them correctly. This leads to metric results that are certain and, hence, accurate.

The frameworks reported in~\cite{jouck_ptandloggenerator:_2016,jouck_simulating_2017} are clearly not the only to generate process models and event logs. While the framework would allow one to plug different model and log generators, the choice has fallen onto those frameworks because they provide an API that allows one to invoke them from code, as our scientific workflow requires. For example, PLG~\cite{DBLP:conf/bpm/Burattin16} only allows a GUI interaction; also, PLG does not support certain patterns, namely long-term dependencies, silent transitions and duplicate activity labels.

Finally, the classification approach of the proposed evaluation framework builds upon established principles and methods from the machine learning domain. See~\cite{japkowicz_evaluating_2011} for more information on the empirical evaluation of learning algorithms using a classification perspective.

\section{Conclusions}
\label{sec:conclusions}

Many process discovery algorithms have been proposed in recent years. As a result the evaluation of process discovery techniques to decide which technique performs best on a given event log has gained importance. However, existing evaluation frameworks have several important drawbacks.

This paper presented a new evaluation framework to overcome existing limitations. The new framework is independent from the discovered model's modeling notation by adopting a classification approach. It starts by defining a population of process models using different behavioral characteristics. From this population a set of models and event logs is randomly sampled. Using a 10-fold cross-validation approach, the event logs are split into training and test logs. Then the framework adds noise to half of the test traces to generate non-fitting traces. The discovery algorithm learns a model based on the training log and classifies the test traces as fitting or non-fitting. The framework then combines the classification results in a confusion matrix together with the metrics recall, precision and $F_1$ score. These metrics are the input for the final statistical tests that are used to determine whether significant differences between algorithms exist or whether certain model or log characteristics have a significant effect on algorithm's performance.

The framework is designed as a scientific workflow. The workflow is then implemented in the RapidMiner tool so that evaluation experiments can be automated, shared between researchers and extended to include new discovery techniques. The framework allows researchers to benchmark discovery algorithms as well as to perform a sensitivity analysis to evaluate whether certain model or log characteristics have a significant effect on algorithm's performance.

The framework has been validated by conducting an extensive experiment involving four process discovery algorithms, five control-flow characteristics and two levels of infrequent behavior. The experiment has shown the usefulness and flexibility of the framework. Additionally, future research opportunities were identified.


%

%

\begin{IEEEbiography}[{\includegraphics[width=1in,height=1.25in,clip,keepaspectratio]{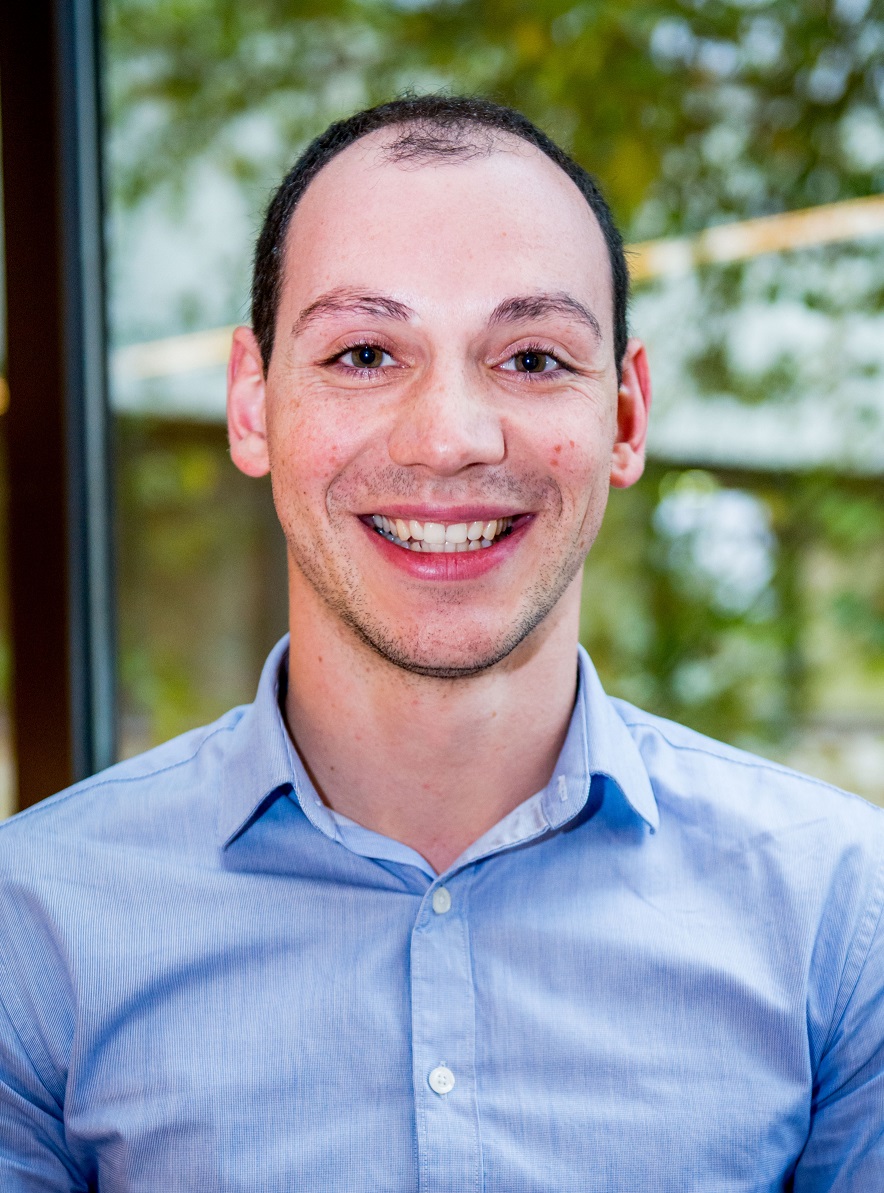}}]{Toon Jouck}
received the M.Sc. degree Business and Information Systems Engineering at Hasselt University, Belgium. He is currently a Ph.D. candidate in the Department of Business Informatics, Hasselt University, and his current research interests include business process mining and management, data analytics and experimental design.
\end{IEEEbiography}

\begin{IEEEbiography}[{\includegraphics[width=1in,height=1.25in,clip,keepaspectratio]{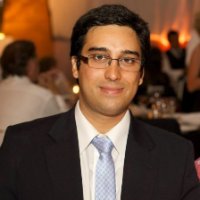}}]{Alfredo Bolt}
received the M.Sc. degree in Computer Science at Pontificia Universidad Catolica de Chile, Chile. He is currently a Ph.D. candidate in the Department of Mathematics and Computer Science at the Technische Universiteit Eindhoven (TU/e), the Netherlands. His current research interests include process cubes, process comparison and process mining workflows.
\end{IEEEbiography}

\begin{IEEEbiography}[{\includegraphics[width=1in,height=1.25in,clip,keepaspectratio]{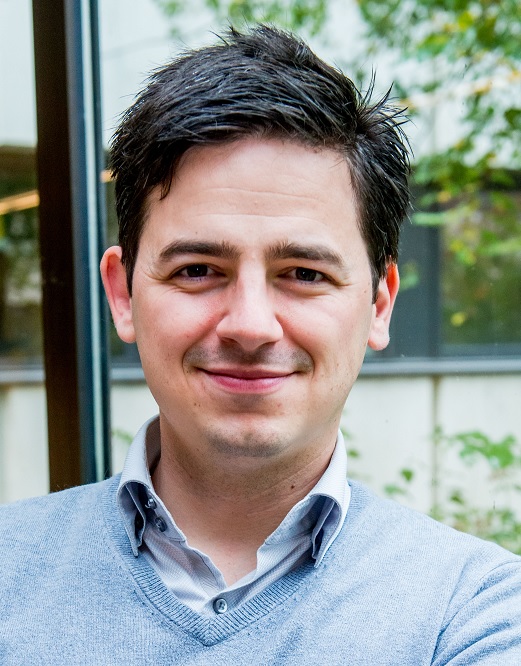}}]{Beno\^it Depaire}
is an Associate Professor at Hasselt University, Belgium. He has authored and co-authored numerous papers published in international journals and conference proceedings. His current research interests include business process management, experimental design, process simulation and IT-business models. His aim is to gain empirically-validated insights from process-related data to understand and improve business processes.
\end{IEEEbiography}

\begin{IEEEbiography}[{\includegraphics[width=1in,height=1.25in,clip,keepaspectratio]{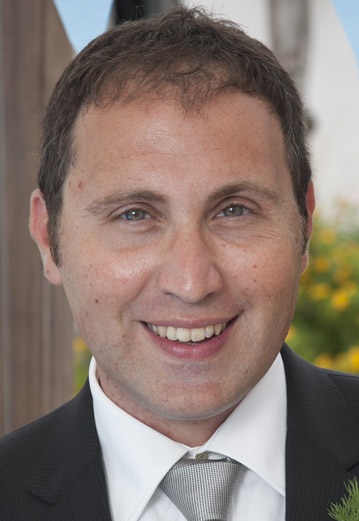}}]{Massimiliano de Leoni}
Massimiliano de Leoni is an Assistant Professor of Computer Science at the Technische Universiteit Eindhoven (TU/e), the Netherlands.
He is constantly publishing papers for the major conference venues and high-reputation journals in  Business Process Management and Information Systems.
His research interests are in the area of Process-aware Information Systems and Business Process Management, predominantly focusing on multi-perspective process mining, process-aware decision support systems and OLAP-based process-mining techniques.
\end{IEEEbiography}

\begin{IEEEbiography}[{\includegraphics[width=1in,height=1.25in,clip,keepaspectratio]{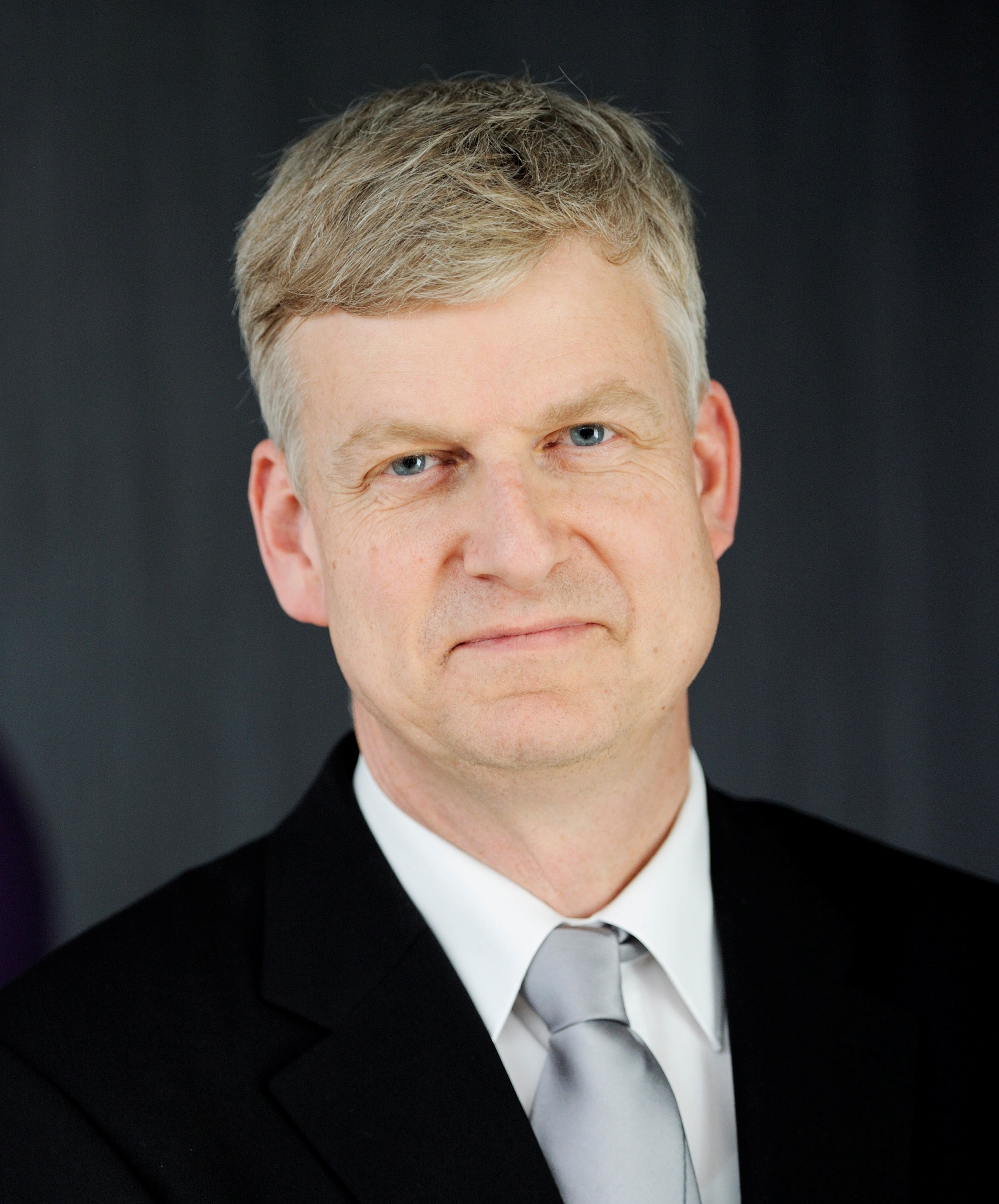}}]{Wil van der Aalst}
Prof.dr.ir. Wil van der Aalst is a professor at Aachen.  His research interests include process mining, Petri nets, business process management, workflow management, process modeling, and process analysis. Wil van der Aalst has published over 200 journal papers, 20 books (as author or editor), 450 refereed conference/workshop publications, and 65 book chapters.  He is also an elected member of the Royal Netherlands Academy of Arts and Sciences, the Royal Holland Society of Sciences and Humanities, and the Academy of Europe.
\end{IEEEbiography}

\end{document}